\documentclass[10pt,journal]{IEEEtran}
\usepackage[table]{xcolor}
\usepackage{cite}      
\usepackage{graphicx} 
\usepackage{amsmath}
\usepackage{fourier}
\usepackage[switch]{lineno}

\usepackage{psfrag}    
\usepackage{subfigure}
\usepackage{hyperref}
\usepackage{cleveref}
\usepackage{placeins}
\usepackage{multirow}
\usepackage{titlesec}
\titlespacing{\section}{0pc}{1pc}{0pc}
\titlespacing{\subsection}{0pc}{1pc}{0pc}
\usepackage{multicol}
\usepackage{lipsum}
\usepackage{url}
\hypersetup{
    colorlinks = true,
    linkcolor = black,
    citecolor= black,
    urlcolor =blue
  }
\usepackage{stfloats}  
\usepackage{amsmath}   
\usepackage{titlesec}
\titlespacing\section{0pt}{12pt plus 4pt minus 2pt}{0pt plus 2pt minus 2pt}
\titlespacing\subsection{0pt}{12pt plus 4pt minus 2pt}{0pt plus 2pt minus 2pt}
\titlespacing\subsubsection{0pt}{12pt plus 4pt minus 2pt}{0pt plus 2pt minus 2pt}

\usepackage{array}
\usepackage[rawfloats=true]{floatrow} 
\restylefloat{figure} 
\usepackage[font=small,skip=0pt]{caption}
\newcolumntype{M}[1]{>{\centering\arraybackslash}m{#1}}
\begin{document}
\title{SimHumalator: An Open Source WiFi Based Passive Radar Human Simulator For Activity Recognition}
\author{
    \IEEEauthorblockN{Shelly Vishwakarma\IEEEauthorrefmark{1}, Wenda Li\IEEEauthorrefmark{1}, Chong Tang\IEEEauthorrefmark{1}, Karl Woodbridge\IEEEauthorrefmark{2}, Raviraj Adve\IEEEauthorrefmark{3},  Kevin Chetty\IEEEauthorrefmark{1}}
    \IEEEauthorblockA{\IEEEauthorrefmark{1}Department of Security and Crime Science, University College London, UK}
    \IEEEauthorblockA{\IEEEauthorrefmark{2}Department of Electronic and Electrical Engineering, University College London, UK} 
    \IEEEauthorblockA{\IEEEauthorrefmark{3} Department of  Electrical and Computer Engineering, University of Toronto, Canada} 
    \{s.vishwakarma, wenda.li,chong.tang.18, k.woodbridge,  k.chetty\}@ucl.ac.uk, 	
rsadve@comm.utoronto.ca
}
\maketitle

\begin{abstract}
This work presents a simulation framework to generate human micro-Dopplers in WiFi based passive radar scenarios, wherein we simulate IEEE 802.11g complaint WiFi transmissions using MATLAB's WLAN toolbox and human animation models derived from a marker-based motion capture system. We integrate WiFi transmission signals with the human animation data to generate the micro-Doppler features that incorporate the diversity of human motion characteristics, and the sensor parameters. In this paper, we consider five human activities. We uniformly benchmark the classification performance of multiple machine learning and deep learning models against a common dataset. Further, we validate the classification performance using the real radar data captured simultaneously with the motion capture system. We present experimental results using simulations and measurements demonstrating good classification accuracy of $\geq$ 95\% and $\approx$ 90\%, respectively.   \end{abstract}
\providecommand{\keywords}[1]{\textbf{\emph{Keywords--}}#1}
\begin{IEEEkeywords}
Passive WiFi Sensing, micro-Dopplers, activity recognition, deep learning, simulator
\end{IEEEkeywords}

\IEEEpeerreviewmaketitle

\section{Introduction}
\label{Sec:Intro}
Humans are non-rigid bodies whose motion when illuminated by radio-frequency (RF) fields, gives rise to frequency modulations, popularly known as micro-Dopplers. Over the last decade, radar sensors have used these micro-Doppler signatures to detect, track and classify human activities for numerous applications ranging from law enforcement, security, and surveillance purposes \cite{fioranelli2015classification,fioranelli2015aspect,chen2019dopnet,fairchild2013micro,fairchild2014classification}
to various ubiquitous sensing applications such as assisted living for the elderly e.g., fall detection \cite{erol2018radar,seifert2019toward,le2019radar,hall2019abnormal,erol2020motion}, bio-medical applications for non-intrusively monitoring patients \cite{chen2016activity,seyfiouglu2017deep,fioranelli2019radar,shah2019rf}, and smart home applications such as occupancy detection \cite{amin2019rf,sevegi2019} and hand gesture recognition \cite{kim2016hand,li2017sparsity}. Micro-Dopplers have been observed with active and passive radar sensors \cite{chen2003analysis,chen2006micro,chetty2011through,vishwakarma2017detection,chen2019micro}. However, in recent years, due to the rise in transmitters of opportunity, passive sensing has attracted significant attention for indoor monitoring applications, \cite{chetty2011through,li2018wifi,tan2018exploiting,li2020passive,chen2020respiration}. Passive sensing leverages the existing RF signals and infrastructure; its receiver only nature leads to low power consumption and less complex systems. In this paper, we consider a passive radar exploiting WiFi transmitters.  

Passive WiFi radar (PWR) data have been gathered through actual measurements in laboratory conditions using various radar hardware platforms \cite{li2017passive,li2020passive,tan2018exploiting,shah2019rf}. The use of measurement data is important to perform a thorough evaluation of various signal processing and machine learning algorithms' performances in realistic scenarios. However, the performances of these algorithms are generally tied to the large volumes of high-quality training data. Unlike the fields of vision and image processing, the radar community has limited access to databases that contain large volumes of experimental data. Therefore, it would be beneficial to have some means of simulating radar returns in passive WiFi sensing scenarios that would serve as a means of generating large volumes of training data. The simulation data can be used for preliminary evaluation of different algorithms, and studying effects of radar phenomenology. 

There exist multiple methods to simulate human micro-Doppler data in active radar scenarios. The earliest method modeled the human leg as a double pendulum structure \cite{smith2008multistatic}. However, this model does not simulate radar returns from other human body parts such as the torso and arms, which also contribute significantly to the micro-Doppler returns. The second method developed a human walking model based on extensive biomechanical experiments \cite{boulic1990global,chen2019micro}. Here, twelve analytical expressions govern the motion trajectories of 17 reference points on the human body as a function of the human's height and the relative velocity. This approach is based on a constant velocity model. Therefore, it cannot capture variations in more complex motions such as falling, sitting, jumping. 

The third technique uses animation data from motion capture systems to model more realistic and complex human motions. There are two types of motion capture technology available- marker-based and marker-less. in a marker-based motion capture system, several markers are placed on the live actor's body parts such as head, torso, arms, and legs to capture their three-dimensional time-varying positions in space. The authors of \cite{ram2008simulation,ssramthroughwall}, first developed a complete end-to-end active radar simulator of humans using a marker-based motion capture technique. The radar scatterings were simulated by integrating the animation data of humans with primitive shapes based electromagnetic modeling. Alternatively, the authors of \cite{erol2015kinect,singh2018simulation}, gathered animation data using a 
marker-less motion capture technology based on Microsoft's Kinect. We adopt a simulation methodology similar to that presented in \cite{ram2008simulation} (for markers-based),\cite{singh2018simulation} (is for Kinect) but, for passive sensing. 

Passive WiFi sensing identifies target reflections through cross-correlation based processing to determine range and Doppler information. We simulate the IEEE 802.11g standard WiFi transmissions using MATLAB's WLAN toolbox \cite{matlab2018toolbox}, and human animations using a marker-based motion capture system called Phase-Space \cite{MoCap_2020}. The simulator generates the micro-Doppler radar returns as a function of-  \textbf{target motion characteristics} (aspect of the target to the radar, initial position in space, different motions type), \textbf{sensor parameters} (different PWR radar configurations-monostatic, bistatic-in line, and bistatic circular, waveforms) and \textbf{radar signal processing parameters} (such as coherent processing interval (CPI), and pulse repetition interval (PRI)). By varying these parameters and PWR radar operating conditions, we are able to generate a vast simulation database. We benchmark the performances of different machine learning and deep learning classification algorithms against this common dataset. 

There is currently no open-source simulation tool for generating human micro-Doppler radar data to the best of our knowledge. Therefore, we have publically released the simulator to assist the radar and sensing communities to benchmark future algorithms and generate large volumes of high quality and diverse radar datasets. We also believe that using our simulator will reduce the expense and labour involved in data acquisition by other researchers. The simulator is available for the interested users on \url{https://uwsl.co.uk/}. 

To validate the performance of our simulator, a series of experiments were carried out in which both radar returns from human targets, and motion capture (MoCap) data from the PhaseSpace system were recorded simultaneously. We considered five human motion classes- human standing in a fixed position whilst rotating their body (HBR), human kicking (HK), human punching (HP), human grabbing an object (HG), and a human walking back and forth in front of the radar (HW). The micro-Doppler signatures corresponding to each of these activities share standard features because of the motion similarity. All these motions are periodic and thus have alternating positive and negative micro-Doppler features. Therefore, it becomes challenging for any classifier to discern the correct motion class. We evaluate different classical machine learning (handpicked features, cadence velocity features, sparse features) \cite{kim2009human,ricci2015recognition,li2017sparsity,vishwakarma2017dictionary,vishwakarma2018dictionary}, and deep learning-based classifiers (deep convolutional neural network, AlexNet, GoogLeNet, and ResNet18) \cite{iandola2016squeezenet,zhong2015high,ayyachamy2019medical}, using both the simulation and measurement data for these five motion categories. We observe average classification accuracy $\geq$ 90 in almost all the deep learning frameworks. 

To summarize, our contributions in this paper are the following: 
\begin{enumerate}
    \item The public release of a PWR human simulator that can simulate radar returns as a function of target motion characteristics, sensor parameters, and radar signal processing parameters. The Simulator incorporates motion capture data for five human motion classes: human body rotating (HBR), human Kicking (HK), human Punching (HP), human grabbing an object (HG), and human Walking (HW).
    \item Uniform benchmarking across different classifiers: traditional machine learning (handpicked features + support vector machines (SVM), automatic feature extraction + SVM), deep neural networks (both pre-trained and untrained)
    \item Uniform benchmarking across different PWR geometries:
     monostatic with a single aspect angle, monostatic with aspect angles variation, bistatic with bistatic angle variation.  
    \item Performance validation using measured PWR data. 
\end{enumerate}

Our paper is organized as follows. Section \ref{sec:simulations} describes our simulation framework and simulation database generation. Next, Section \ref{sec:simulation_results} presents the classification results of different classification algorithms in different PWR scenarios. We describe our measurement data collection and validation of all the algorithms' performance in Section \ref{sec:measurements}. We finally conclude our paper in Section \ref{sec:conclusion}. 

\section{SimHumLator}
\label{sec:simulations}
In this section, we first describe the simulation framework used for developing the PWR human simulator. We give a brief introduction of the simulator's capability in generating a diverse set of human micro-Doppler signatures for different- radar parameters, target parameters, and different radar signal processing parameters. Finally, we present the simulation dataset generated using the tool. 

\subsection{Simulation Framework}
\begin{figure*}[t]
\centering
\includegraphics[scale=0.60]{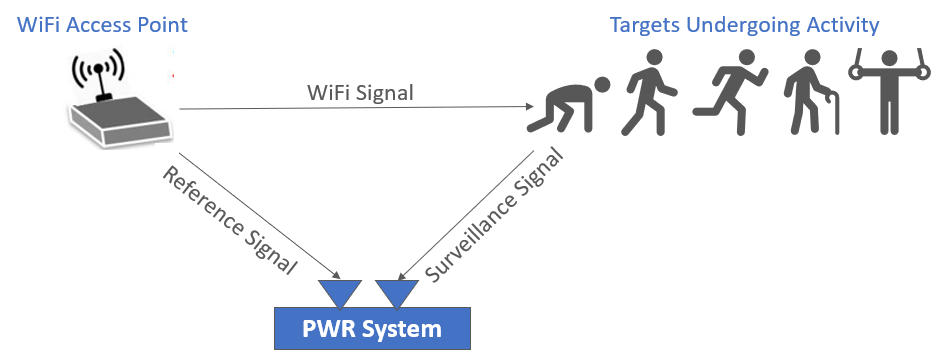}
\caption{A typical passive WiFi radar scenario comprising of transmissions from WiFi access points and targets undergoing motion in the same propagation environment}
\label{fig:PWR_setup}
\end{figure*}

A typical PWR sensing setup is shown in Fig.~\ref{fig:PWR_setup}. It comprises of two antennas- reference and surveillance antenna and a signal processing unit. The reference antenna is a directional antenna that captures the direct signal from the WiFi access point (AP). On the other hand, the surveillance antenna is omni-directional to capture the reflected signals of the human targets present anywhere in the sensing area. The signals reflected off the targets are time-delayed, and Doppler-shifted direct signals. The time delay is directly proportional to the target range, Doppler shift to the target's velocity, and the complex reflectivity to the target's size, shape, and material. The radar signal processing unit aims to estimate these parameters using both the direct and the reflected signals. It employs match-filtering in the digital domain wherein the direct and the reflected signals are cross-correlated in the delay-Doppler plane to generate plots of the cross ambiguity function (CAF). Match-filtering is adopted to maximize the signal-to-noise ratios.  

\begin{figure}[t]
\centering
\includegraphics[scale=0.5]{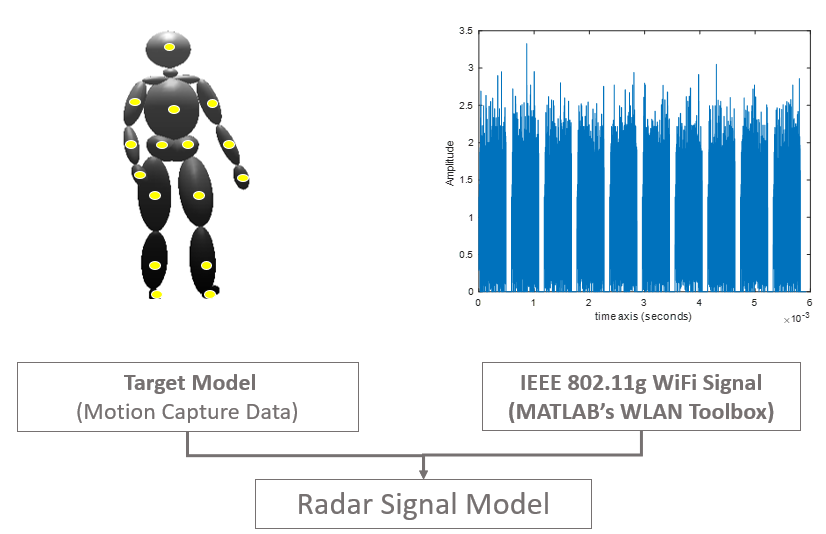}
\caption{Radar signal model after integration of the target model with WiFi transmissions}
\label{fig:Hybrid_model}
\end{figure}

We simulate a standard IEEE 802.11g WiFi signal using MATLAB's WLAN toolbox and human animation data using a marker-based motion capture technology. We integrate both together to generate the radar scatterings off the humans, as shown in Fig.~\ref{fig:Hybrid_model}. We describe these steps in greater detail in the following sections.

\subsubsection{PWR Signal Model}
\begin{figure}[t]
\centering
\includegraphics[scale=0.4]{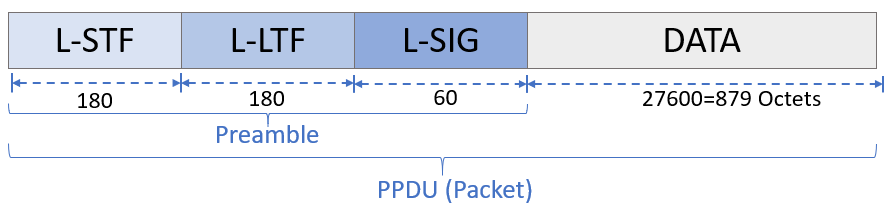}
\caption{IEEE 802.11g standard complaint OFDM transmission packet structure}
\label{fig:PPDU}
\end{figure}

We use MATLAB's WLAN toolbox to generate IEEE 802.11g standard-compliant orthogonal frequency-division multiplexing (OFDM) waveforms \cite{matlab2018toolbox}. The other standards that can be synthesized with the tool are 802.11a, b, g, n, ad, ac, ah, and ax. However, we restrict our discussion to 802.11g packet structures to mimic real WiFi transmission formats at the 2.4GHz band with a channel bandwidth of $BW=20MHz$. The physical layer of IEEE 802.11g standards use a packet-based protocol. Each transmission packet (a physical layer conformance procedure (PLCP) protocol data unit (PPDU)) comprises of a preamble and then the data, as shown in Fig.~\ref{fig:PPDU}. 
The preamble field is embedded with three sub-fields each comprising of a number of time-domain samples- 180 samples in legacy short training field (L-STF), 180 samples in legacy long training field (L-LTF), and 60 samples in legacy signal field (L-SIG). L-STF possesses excellent correlation properties and is therefore used to detect the start of the packet, L-LTF field for communication channel estimation, and the third preamble field L-SIG to indicate the amount of data transmitted (in octets). On the other hand, the data field contains information such as user payload, medium access control (MAC) headers, and the cyclic redundancy check (CRC) bits. The data bits together with the preamble bits form a discrete-time sequence $x_T[n]$. We synthesized multiple such sequences to form a continuous stream of WiFi transmission signals. The sequences differ by a delay $T_p$ equivalent to one sequence's transmission time and a short idle time between sequences.

The continuous stream of sequences $x_T[n]$ is then used to create the baseband signal $x_T(t)$ as shown in (\ref{eq:baseband}). 
\begin{equation}
\begin{aligned}
\label{eq:baseband}
x_T(t)&=\frac{1}{\sqrt(P)} \sum_{p=1}^{P} \sum_{n=1}^{N}x_T[n] h_T(t-n t_s-pT_p)
\end{aligned}
\end{equation}
Here $N$, is the total number of time domain samples with each sample of duration $t_s=(1/BW)=50n$sec, $h_T(t)$ is a pulse shaping filter, $P_t$ is the transmit signal power amplification factor, P is the number of transmission packets and $T_p$ corresponds to the pulse repetition interval (PRI). 

The baseband signal is finally up-converted for transmission at a carrier frequency of $f_c=2.4GHz$, as shown in (\ref{eq:passband}). 
\begin{equation}
\label{eq:passband}
y_T(t)=x_T(t)e^{-j2 \pi f_c t}
\end{equation}

\subsubsection{Dynamic Target Model}
\begin{figure*}[t]
\centering
\includegraphics[scale=0.60]{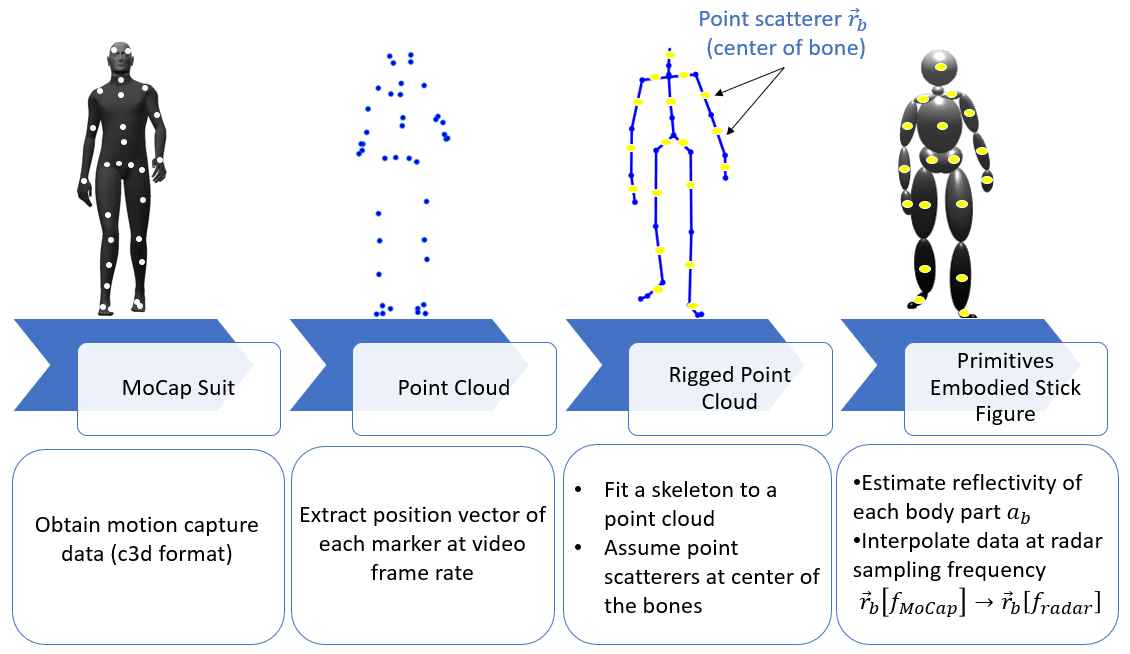}
\caption{Simulation framework for generating an electromagnetic human primitive scattering model from motion capture animation data.}
\label{fig:Mocap_generation}
\end{figure*}

We present a realistic human simulation model in Fig.~\ref{fig:Mocap_generation}. 
The first step is to capture the animation data of dynamic humans. We use an active tracking Phase-Space system to gather the 3-dimensional time-varying location of several LED markers placed on the live actor's bodysuit \cite{MoCap_2020}. The Phase-Space system consists of 8 cameras that can track 71 markers at a frame rate of $960$ frames per second (FPS). However, we use only 25 markers to extract information from 25 joints on the human body. There are several standard formats for saving an animation data file, such as C3D, BVH, ASF/AMC. Our Phase-Space system exports animation data as a set of three-dimensional points into the C3D files. The C3D file format is not a hierarchical format like BVH and ASF/AMC; therefore, we rig the point cloud data with a skeleton model to specify the bones' hierarchical distribution in the human body. We finally integrate the human animation model with an electromagnetic scattering center model. 

We embody the human skeleton with $ B $ elementary shapes to model different parts of the body, such as the- torso, arms, and legs using ellipsoids and the head using a sphere. We assume the radar scattering centers to be lying approximately at the center of these primitive shapes. The complex reflectivities $a_b(t)$ of each of these $B$ primitive shapes depends on various factors such as the material properties, aspect angle $\theta_b(t)$, and the relative position $r_b(t)$, of the scattering center on the primitive shape with respect to the radar \cite{chen2003analysis}.  The reflectivity of a primitive at any time instant $t$ is given by
\begin{equation}
    \label{eq:RCSellipsoid}
    a_b(t) =\frac{\zeta(t) \sqrt{\sigma_b(t)}}{r_b^2(t)}
\end{equation}\normalsize
Here, $\zeta(t)$ subsumes propagation effects such as attenuation, antenna directivity, processing gains, $\sigma_b(t)$ is the radar cross section of the primitives. The RCS of primitive shapes are well characterised at microwave frequencies. The RCS of an ellipsoid of length $L_b$ and radius $R_b$ is given by
\begin{equation}
\sigma_b(t) = \sqrt\Gamma  \frac{\frac{\pi}{4}R_b^4 L_b^2}{R_b^2\sin^2 \theta_b(t)+ \frac{1}{4}L_b^2\cos^2 \theta_b(t)}
\end{equation}\normalsize
We incorporate the effect of the dielectric properties of human skin into the RCS estimation through the Fresnel reflection coefficient $\Gamma$. We assume the human to be a single layer dielectric with a dielectric constant of 80 and conductivity of 2 S/m.

\subsubsection{Hybrid electromagnetic radar scattering from dynamic humans}
The WiFi AP is assumed to transmit P packets in the propagation channel comprising human target with B point scatterers. The received signal $y_R(t)$ comprises both the direct signal and the complex sum of time-varying reflections from each point scatterer. The target reflections are simply attenuated, time-delayed $\tau_b=2r_b/c$, and Doppler-shifted $f_{Db}=2v_b f_c/c$, versions of the transmitted signal. Ignoring multipath, the baseband received signal on the survelliance channel can be represented as  
\begin{equation}
\label{eq:totalRX}
y_R(t)= \sum_{p=1}^{P}\sum_{b=1}^{B}a_b(t) y_T(t-\tau_b-pT_p)e^{-j2 \pi f_{Db} pT_p}+ z(t)
\end{equation}
Here, $c=3 \times 10^8$m/s is the speed of light and $z(t)$ is the additive circular-symmetric white noise. 

Since the WiFi AP transmissions are a continuous stream of signals, the received signal is also a long sequence of data spanning a duration $T_{Total}$. Cross-correlation and Fourier processing over this large set of data is a computationally expensive task. Therefore, we process the received data in $M$ batches each of duration equal to one coherent processing interval of $T_{Total}=MT_{CPI}$. We assume the target Doppler frequencies to remain constant within one CPI $T_{CPI}$. The approximation follows from the fact that maximum Doppler shift $f_{Dmax}$ is always $f_{Dmax}\ll  1/(T_p)$; therefore, the phase rotation can be approximated as a constant over one CPI. The baseband digitized signal for one CPI is shown in \ref{eq:totalRX1}.
\begin{equation}
\label{eq:totalRX1}
y_R[n,p]= \sum_{p=1}^{P}\sum_{b=1}^{B}a_b[n] y_T(nt_s-\tau_b-pT_p)e^{-j2 \pi f_{Db} pT_p}+ z[n]
\end{equation}
Here, $n$ is the index of fast time samples within the $p^th$ PRI, and $p$ is the index of PRIs (slow time samples) in one CPI.  

\subsubsection{CAF processing}
\begin{figure*}[t]
\centering
\includegraphics[scale=0.60]{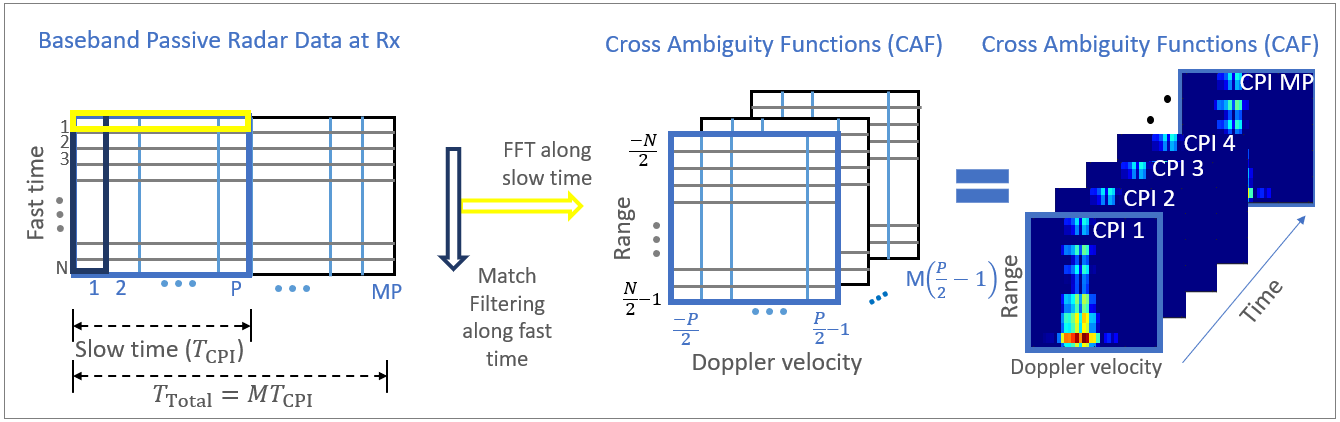}
\caption{cross ambiguity function generation through cross-correlation between the direct WiFi transmission and the reflected signals off the targets}
\label{fig:CAF_gen}
\end{figure*}

We implement cross ambiguity function processing over the two-dimensional received data $y_R[n,p]$ and the direct reference signal data ( that is without target reflections) to compute the delay $\tau_b$ and Doppler information $f_{Db}$ of the target. The adopted CAF processing is shown in Fig.~\ref{fig:CAF_gen}. We perform match-filtering along the fast time samples and fast Fourier transform (FFT) along the slow time samples to generate CAFs for the $m^{th}$ CPI. The CAF processing is implemented as  
\begin{equation}
\label{eq:RD}
\chi_{m}^{RD}[\tau,f_D]= \frac{1}{NP}\sum_{p=1}^{P} x_T(nt_s) y_R^*(nt_s-\tau-pT_p)e^{j2 \pi f_{D} pT_p}
\end{equation}

\begin{figure*}[t]
\centering
\includegraphics[scale=0.55]{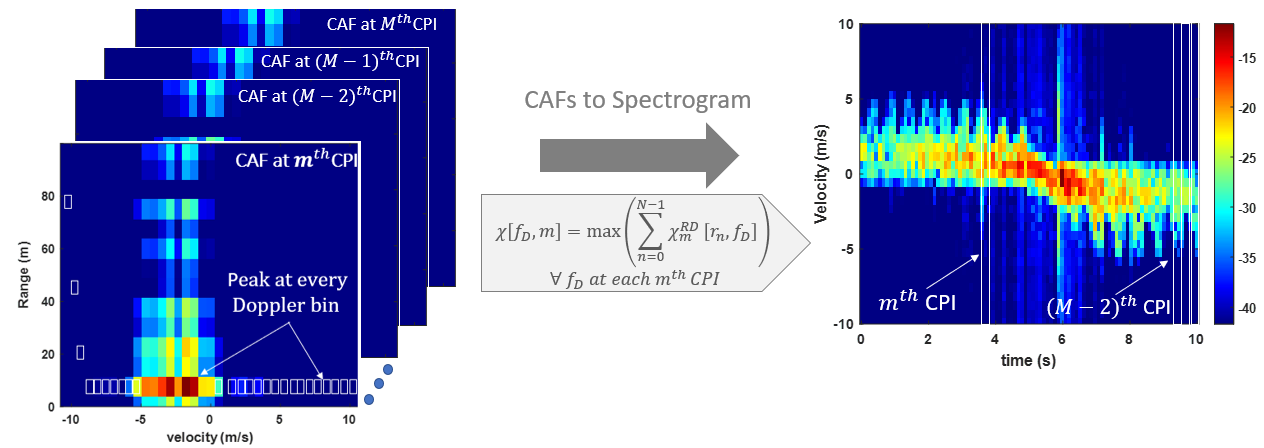}
\caption{Simulation methodology to generate micro-Doppler spectrograms using multiple CAFs spanning the entire duration of motion}
\label{fig:CAF_to_spec}
\end{figure*}

Multiple CAFs spanning a duration of $T_{Total}$, are processed to generate the Doppler-time spectrogram as shown in Fig. \ref{fig:CAF_to_spec}. 
Here, for each CPI, the peaks along the range axis are coherently added for each Doppler bin. Mathematically this procedure can be represented as
\begin{equation}
\label{eq:DT}
\chi^{DT}[f_D,m]=\max_{n}(\sum_{n=1}^{N}\chi_{m}^{RD}[\tau,f_D]) \quad \forall \ f_D \ at \ m^{th} \ CPI
\end{equation}

\subsection{Simulation Database Generation}
The PWR signal parameters we use in this work are provided in TABLE \ref{table:sim_param}. We fix the CPI to 0.3s, which is sufficient to capture time-varying micro-Doppler features with a Doppler resolution of $5$Hz in joint time-frequency space.
\begin{table}[]
\centering
\caption{Simulated PWR Parameters}
\label{table:sim_param}
\begin{tabular}{|c|c|} \hline 
\textbf{Radar Parameters} & \textbf{Values} \\ \hline 
Carrier frequency ($f_c$) & $2.4$GHz \\ \hline 
Bandwidth ($f_s$) & $20$MHz \\ \hline 
Pulse Repetition Frequency ($PRF=1/T_p$) & $500$Hz \\ \hline 
Coherent Processing Interval & $0.3s$ \\ \hline 
Maximum Doppler ($f_{Dmax}=\pm PRF/2$) & $\pm 250$Hz \\ \hline 
Doppler resolution (${\Delta f}_D=1/CPI$) & $5$Hz \\ \hline 
\end{tabular}
\vspace{-4mm}
\end{table}

TABLE \ref{table:Siml_Data_Set_Description} summarizes our entire data simulated using the described methodology and the radar signal model described in TABLE \ref{table:sim_param}. We simulate the human micro-Doppler data for five motion classes- human body rotating (HBR),  human kicking (HK), human punching (HP), human grabbing an object (HG) and a human walking (HW). 

The number of animation data files in HBR, HK, HP, HG, and HW are 10, 20, 20, 20, and 19 respectively, and differ because some of the data files had many missing markers positions, which resulted in insufficient marker information. Therefore, we decided to drop these files and continued with the remaining ones. Note that the human motions in these repeated measurements were unrestricted and, therefore, the micro-Doppler signatures vary due to differences in gait patterns every simulation. The duration of each measurement is 4.5sec. We use a sliding window of duration 1.5sec with an overlapping time of 0.5sec over the entire signature of duration 4.5sec. It results in 9 spectrograms, each of duration 1.5sec from every motion capture file.
\begin{table*}[]
\centering
\caption{Simulation Dataset Description}
\label{table:Siml_Data_Set_Description}
\begin{tabular}{|c|c|c|c|l|c|c|l|}
\hline
\multirow{3}{*}{\textbf{Target Class}} & \multirow{3}{*}{\textbf{\begin{tabular}[c]{@{}c@{}}Number of MoCap \\ Files\end{tabular}}} & \multirow{3}{*}{\textbf{\begin{tabular}[c]{@{}c@{}}Duration of Data\\ (in each file)\end{tabular}}} & \multicolumn{5}{c|}{\textbf{\begin{tabular}[c]{@{}c@{}}Number of Spectrograms  (1.5sec each)\\ In Radar Configuration\end{tabular}}}                                                                                                                                                                                                                    \\ \cline{4-8} 
                                       &                                                                                            &                                                                                                     & \multicolumn{3}{c|}{\textbf{Monostatic}}                                                                                                                                                                                    & \multicolumn{2}{c|}{\textbf{Bistatic}}                                                                                    \\ \cline{4-8} 
                                       &                                                                                            &                                                                                                     & \multicolumn{2}{c|}{\textbf{\begin{tabular}[c]{@{}c@{}}Aspect Angle 0\\ (Overlapping Window 0.5sec)\end{tabular}}} & \textbf{\begin{tabular}[c]{@{}c@{}}Varying Aspect Angle\\ 0:5:360\\ (No Overlapping)\end{tabular}}     & \multicolumn{2}{c|}{\textbf{\begin{tabular}[c]{@{}c@{}}Varying Bistatic Angle\\ 0:5:360\\ (No Overlapping)\end{tabular}}} \\ \hline
HBR                                    & 10                                                                                         & 4.5 sec                                                                                             & \multicolumn{2}{c|}{90}                                                                                           & 2190                                                                                                   & \multicolumn{2}{c|}{2190}                                                                                                 \\ \hline
HK                                     & 20                                                                                         & 4.5 sec                                                                                             & \multicolumn{2}{c|}{180}                                                                                           & 4380                                                                                                   & \multicolumn{2}{c|}{4380}                                                                                                 \\ \hline
HP                                     & 20                                                                                         & 4.5 sec                                                                                             & \multicolumn{2}{c|}{180}                                                                                           & 4380                                                                                                   & \multicolumn{2}{c|}{4380}                                                                                                 \\ \hline
HG                                     & 19                                                                                         & 4.5 sec                                                                                             & \multicolumn{2}{c|}{180}                                                                                           & 4161                                                                                                   & \multicolumn{2}{c|}{4161}                                                                                                 \\ \hline
HW                                     & 20                                                                                         & 4.5 sec                                                                                             & \multicolumn{2}{c|}{171}                                                                                           & 4380                                                                                                   & \multicolumn{2}{c|}{4380}                                                                                                 \\ \hline
\multicolumn{3}{|c|}{\textbf{Total Data in Each Configuration}}                                                                                                                                                                           & \multicolumn{2}{c|}{\textbf{801}}                                                                                  & \textbf{19491}                                                                                         & \multicolumn{2}{c|}{\textbf{19491}}                                                                                       \\ \hline
\end{tabular}
\end{table*}

We generate the spectrograms in three different PWR radar configurations:
\begin{enumerate}
    \item \textbf{PWR Monostatic Configuration With Fixed Target Aspect Angle}:\\
    In this radar configuration, the WiFi AP and radar receiver are co-located, and the target moves at $0^{\circ}$  aspect angle with respect to the radar receiver that is in direct line-of-sight with respect to the receiver. 
    The micro-Doppler signatures for this configuration are shown in Fig.~\ref{fig:simul_spec}. Each spectrogram is 1.5sec long. 
    
    Fig.~\ref{fig:simul_spec}(a), presents the spectrogram of a human undergoing a body rotation motion. We can observe both positive and negative Dopplers due to the rotational motion of the body. Since the human is standing in a place while performing the motion, and there is no bulk body translational motion, the body Doppler is mostly centered around zero. 
    
    This observation holds even for three other motion categories- human kicking, human punching, and human grabbing, as shown in Fig.~\ref{fig:simul_spec}(b)(c) and (d), respectively. We notice that there are only minor differences in the micro-Doppler patterns within these motion classes due to the similarity of motions. 
    
    The fifth target class shown in Fig.\ref{fig:simul_spec}(e) corresponds to a human walking in front of the radar. The human always walks in the direction of the radar, thus resulting in mostly positive micro-Dopplers. 
    \begin{figure*}[t]
    \centering
    \includegraphics[scale=0.60]{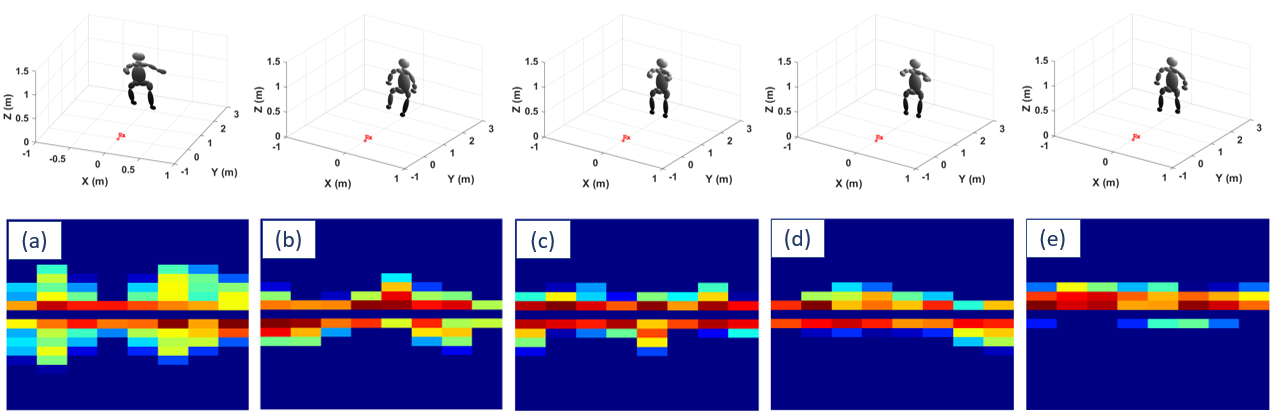}
    \caption{Radar micro-Doppler signatures for a human undergoing (a) a body rotation motion, (b) kicking motion, (c) punching motion, (d) grabbing an object motion and (e) walking in the direction of the monostatic configuration of PWR radar.}
    \label{fig:simul_spec}
    \end{figure*}
     We obtain 90, 180, 180, 180, and 171 spectrograms from HBR, HK, HP, HG, and HW motion categories. 
    \item \textbf{PWR Monostatic Configuration For Varying Target Aspect Angles}\\
    In most realistic scenarios, the human motions might not be restricted to a single aspect angle with respect to the radar. In such scenarios, the spectrograms might differ significantly. It could be due to the shadowing of some part of the human body if captured at different angles. 
    
    Fig.\ref{fig:aspect_bistatic_angle_var}(a)-(d), shows the spectrograms of a human walking at four different aspect angles- $0^{\circ}$, $60^{\circ}$, $120^{\circ}$ and $180^{\circ}$ with respect to the radar respectively. The signatures at $60^{\circ}$ shown in Fig.\ref{fig:aspect_bistatic_angle_var}(b), mostly have reduced positive Dopplers compared to $0^{\circ}$ aspect angle. This is because the target is still approaching but with a reduced radial component towards the radar. The Dopplers become negative when the target aspect angle is $120^{\circ}$. At this angle, the target begins to move away from the radar. Finally, the signatures at $180^{\circ}$ represent a human walking away from the radar, thus have all the negative Dopplers. 
    \begin{figure*}[t]
    \centering
    \includegraphics[scale=0.65]{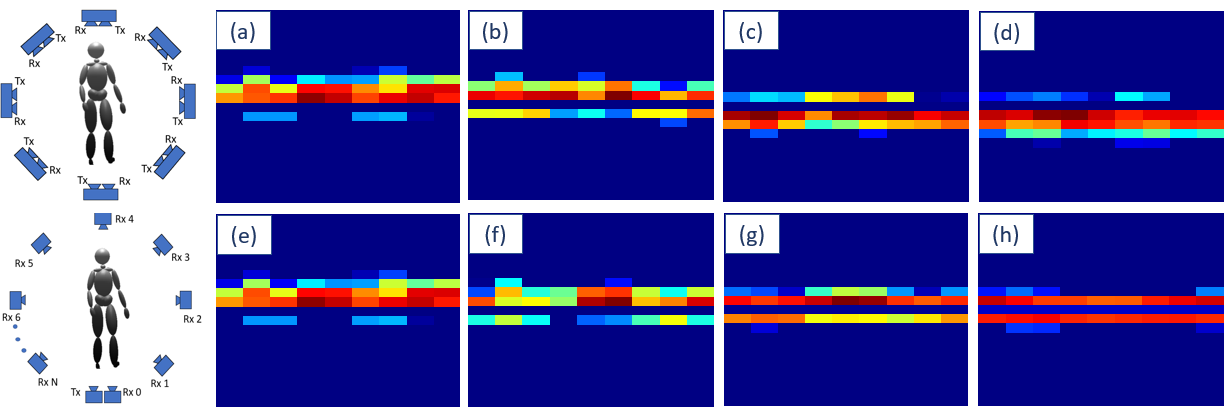}
    \caption{ (a)-(d) Radar micro-Doppler signatures of human walking at four aspect angles $0^{\circ}$, $60^{\circ}$, $120^{\circ}$, and $180^{\circ}$ with respect to the radar respectively. (e)-(h) Radar micro-Doppler signatures of human walking at aspect angles $0^{\circ}$ to the radar at four bistatic radar configurations with following bistatic angle $0^{\circ}$, $60^{\circ}$, $120^{\circ}$ and $180^{\circ}$ respectively.}
    \label{fig:aspect_bistatic_angle_var}
    \end{figure*}

    We gather micro-Doppler data for aspect angles varying from $0^{\circ}$ to $360^{\circ}$ with an interval of $5^{\circ}$ resulting in 73 unique spectrograms of duration 4.5sec each for every file. We further divide the resulting spectrograms into three of duration 1.5sec each. We repeat the process for all of the files and all the target classes. The resulting number of spectrograms are shown in TABLE \ref{table:Siml_Data_Set_Description}.
    
    \item \textbf{PWR Bistatic Configuration For Varying Bistatic Angle And Fixed Target Aspect Angle}\\
    In most passive WiFi sensing scenarios, a certain distance separates the WiFi AP and the radar receiver. Therefore, this motivates us to perform simulations that can capture the effect of varying bistatic angles. In our simulations, we varied the bistatic angle from $0^{\circ}$ to $360^{\circ}$ with an interval of $5^{\circ}$ like the previous case and obtained a very large amount of radar data. The dataset size is shown in TABLE \ref{table:Siml_Data_Set_Description}.
    
    The micro-Doppler signatures of a walking human at four bistatic angles $0^{\circ}$, $60^{\circ}$, $120^{\circ}$ and $180^{\circ}$ are shown in Fig.\ref{fig:aspect_bistatic_angle_var}(e)-(h) respectively. In bistatic scenarios, the Doppler frequency $f_D=\frac{2vf_c}{c}cos(\theta)cos(\beta/2)$, is governed by aspect angle $\theta$ and the bistatic angle $\beta$. For simplicity, we keep the aspect angle of the target fixed at $0^{\circ}$; therefore, the Dopplers only depend upon the bistatic angles $f_D=\frac{2vf_c}{c}cos(\beta/2)$. When $\beta=180^{\circ}$, the forward scatterer position, the Doppler should be zero ($f_D=0$). However, since humans are extended targets, we get some micro-Doppler returns due to the swinging motion of arms and legs. This is evident in Fig.\ref{fig:aspect_bistatic_angle_var}(h). 
\end{enumerate}

\section{Simulation Results}
\label{sec:simulation_results}
In this section, we study the different classification algorithms' robustness to classify micro-Dopplers in more complex scenarios, such as varying aspect angles and varying bistatic angles. We use handpicked features \cite{kim2009human}, Cadence velocity features (CVD) \cite{ricci2015recognition}, and automatically extracted sparse features \cite{li2017sparsity,vishwakarma2018dictionary} from the micro-Doppler signatures to test the performance of classical machine learning-based support vector machine (SVM) classifier \cite{keerthi2000fast}. We then compare their performances with a deep convolutional neural network (DCNN) that has a joint feature extraction and classification framework within the same network. We designed a 24-layered deep neural network comprising three components (convolution layer, pooling layer, and activation functions). We also test some of the pre-trained deep neural networks such as AlexNet, GoogLeNet, and ResNet18. We used 70\% of each target's spectrograms as the training data set, 15\%  as the validation set, and the remaining 15\% as the test data set. The algorithms are run on an Intel(R) Core(TM) i7-5500U CPU running at 2:40 GHz; 16-GB RAM,Windows 10 (64 bit). 

To give readers a better understanding of the sensitivity of the
algorithm's performance to simulation database; we considered the following three classification scenarios.

\begin{itemize}
    \item \textbf{Case 1a: Train using data from a fixed zero aspect angle:} We trained and tested the algorithms' performances using a simulation database generated for a fixed $0{\circ}$ aspect angle of the target. Note that the five target classes considered in the study share common features in micro-Doppler feature space because of the proximity between different motion categories. The resulting classification accuracies are presented in TABLE\ref{table:mono_class_sim}. We observe that all deep neural networks outperform the classical machine learning-based methods and achieve an average classification accuracy of $\approx 99\%$. This is because the deep neural networks being a cascaded structure of neurons, can learn any complex function to create a decision boundary even for nonlinear data considered in the study. The classical machine learning algorithms, on the other hand, are not capable of learning these complex discerning boundaries, resulting in poor classification performances. Classification accuracy using the handpicked features is $88\%$, while the CVD features' performance is $93.4\%$. The performance using sparse features shows that the sparsity-based algorithms can extract underlying features in different hyper-planes, resulting in good classification accuracy of $96.4\%$ even when the motion classes are similar. We also observe that HW class is rarely confused with other target classes since HW is less similar to other motion classes such as HBR, HK, HP, and HG. 
    \begin{table*}[htbp]
    \centering
    \caption{Classification accuracies of multiple algorithms for a simulation database (captured for a fixed aspect angle of the target)}
    \label{table:mono_class_sim}
    \begin{tabular}{|c|c|c|c|c|c|c|c|}
    \hline
    \multirow{2}{*}{\textbf{Target Class/ Algorithm}} & \multirow{2}{*}{\textbf{\begin{tabular}[c]{@{}c@{}}Handpicked\\  Features\end{tabular}}} & \multirow{2}{*}{\textbf{\begin{tabular}[c]{@{}c@{}}CVD \\ Features\end{tabular}}} & \multirow{2}{*}{\textbf{\begin{tabular}[c]{@{}c@{}}Sparse\\  Features\end{tabular}}} & \multicolumn{4}{c|}{\textbf{Neural Networks}}                             \\ \cline{5-8} 
                                                      &                                                                                          &                                                                                   &                                                                                      & \textbf{DCNN} & \textbf{AlexNet} & \textbf{GoogLeNet} & \textbf{ResNet18} \\ \hline
    HBR                                               & 100                                                                                    & 98.9                                                                              & 99.1                                                                                 & 100           & 100              & 100                & 100               \\ \hline
    HK                                                & 83.9                                                                                    & 83                                                                              & 93.1                                                                                 & 100           & 100              & 100               & 96.3              \\ \hline
    HP                                                & 73.8                                                                                     & 92.3                                                                              & 92.4                                                                                 & 100           & 96.3             & 96.3               & 100               \\ \hline
    HG                                                & 82.5                                                                                   & 98.3                                                                              & 97.6                                                                                  & 96.3          & 100              & 100                & 100               \\ \hline
    HW                                                & 100                                                                                      & 94.5                                                                              & 100                                                                                 & 100           & 100              & 100                & 100               \\ \hline
    \textbf{Overall Accuracy (\%)}         & \textbf{88}                                                                                     & \textbf{93.4}                                                                             & \textbf{96.4}                                                                                & \textbf{99.3}          & \textbf{99.3}             & \textbf{99.3}              & \textbf{99.3}             \\ \hline
    \end{tabular}
    \end{table*}
    \item \textbf{Case 1b: Train using data from multiple aspect angles:}Next, we analyse the performance when the algorithms are trained using a simulation database comprising micro-Doppler signatures captured at multiple aspect angles. It is a significantly more challenging and realistic scenario since no aspect angle information is available during the test phase. Note that the spectrograms used in testing have not been used during training. 
    
    TABLE \ref{table:mono_vary_class_sim} presents the resulting classification accuracies across different algorithms. Here, we present the classification accuracy in the form of confusion matrices to draw a more intuitive sense of the results. The average classification accuracies for handpicked features, and CVD features, are 69.2\% and 80.9\%, respectively. The reason for poor performance is that the Doppler spectrogram for a particular motion class at certain aspect angles might look similar to the spectrogram for other motion classes at the same angle. Therefore, the handpicked and the CVD features, are not discriminative enough and result in poor performance. There is a further degradation of 5\% in the performance of sparsity based algorithm. This could be due to sharing of same subspace in sparse domain between different classes for different aspect angles. HW remains the best recognised class amongst all the classes considered in the study as it has more distinctive spectrograms from rest of the classes. On the other hand, algorithms based on deep networks perform exceptionally well even under diverse training and test datasets. It indicates that these algorithms are specifically suited for problems dealing with a great deal of diversity in the radar data. The best performing network is ResNet18, with an average classification accuracy of 97.8\%. 
    
    \begin{table*}[htbp]
    \centering
    \caption{Classification accuracies for a simulation database (captured for varying aspect angle of the target with respect to the radar receiver)}
    \label{table:mono_vary_class_sim}             
    \begin{tabular}{|c|l|cccccc|}
    \hline
    \multicolumn{2}{|c|}{\textbf{Algorithm}}                                                                       & \multicolumn{1}{c|}{\textbf{True/Pred Class}} & \textbf{HBR}  & \textbf{HK}   & \textbf{HP}   & \textbf{HG}   & \textbf{HW}   \\ \hline \hline
    \multicolumn{2}{|c|}{\multirow{5}{*}{\textbf{\begin{tabular}[c]{@{}c@{}}Handpicked \\ Features\end{tabular}}}} & \textbf{HBR}                                  & \textbf{95.4} & 3.4           & 0.9           &  0.2           & 0             \\
    \multicolumn{2}{|c|}{}                                                                                         & \textbf{HK}                                   &  2.4           & \textbf{ 70.9} & 8.7           & 14.5           & 3.5          \\
    \multicolumn{2}{|c|}{}                                                                                         & \textbf{HP}                                   & 2.7             & 32.2         & \textbf{20.1}  & 43          & 1.9            \\
    \multicolumn{2}{|c|}{}                                                                                         & \textbf{HG}                                   & 0.6           & 18.4           & 11.9           & \textbf{68.9} & 0.1           \\
    \multicolumn{2}{|c|}{}                                                                                         & \textbf{HW}                                   & 0.6           & 6.2           & 1.3            & 1.4           & \textbf{90.3} \\ \hline \hline
    \multicolumn{2}{|c|}{\multirow{5}{*}{\textbf{CVD Features}}}                                                   & \textbf{HBR}                                  & \textbf{93.4} & 6.2           & 0.2           & 0.2           & 0             \\
    \multicolumn{2}{|c|}{}                                                                                         & \textbf{HK}                                   & 3.4             & \textbf{57.4}   & 9.7          & 20.9            & 8.6          \\
    \multicolumn{2}{|c|}{}                                                                                         & \textbf{HP}                                   & 1.4             & 7.1           & \textbf{87.4} & 1           & 2.9           \\
    \multicolumn{2}{|c|}{}                                                                                         & \textbf{HG}                                   & 0.1          & 5.7           & 3.8           & \textbf{79.7} & 10.7          \\
    \multicolumn{2}{|c|}{}                                                                                         & \textbf{HW}                                   & 0.8           & 0.8           & 4.4         & 11.3          & \textbf{82.6}   \\ \hline \hline
    \multicolumn{2}{|c|}{\multirow{5}{*}{\textbf{Sparse Features}}}                                                & \textbf{HBR}                                  & \textbf{98.6} & 0.9           & 0             & 0.5           & 0             \\
    \multicolumn{2}{|c|}{}                                                                                         & \textbf{HK}                                   & 0.1           & \textbf{81.8} & 10            & 5.2           & 2.9           \\
    \multicolumn{2}{|c|}{}                                                                                         & \textbf{HP}                                   & 0.6           & 8.1           & \textbf{88.1} & 3.2           & 0             \\
    \multicolumn{2}{|c|}{}                                                                                         & \textbf{HG}                                   & 0             & 3.3           & 6.5           & \textbf{90.2} & 0             \\
    \multicolumn{2}{|c|}{}                                                                                         & \textbf{HW}                                   & 0             & 2.8           & 0.6           & 0.1           & \textbf{96.5} \\ \hline \hline
    \multirow{20}{*}{\textbf{Neural Networks}}                & \multirow{5}{*}{\textbf{DCNN}}                     & \textbf{HBR}                                  & \textbf{100}  & 0             & 0             & 0             & 0             \\
                                                              &                                                    & \textbf{HK}                                   & 0.3           & \textbf{94.2} & 4.4           & 0.8           & 0.3           \\
                                                              &                                                    & \textbf{HP}                                   & 0.9           & 3.3           & \textbf{95.7} & 0             & 0             \\
                                                              &                                                    & \textbf{HG}                                   & 0.3           & 2.9           & 3.4           & \textbf{92.5} & 0.9           \\
                                                              &                                                    & \textbf{HW}                                   & 0.3           & 11.9          & 0             & 0             & \textbf{87.8} \\ \cline{2-8} 
                                                              & \multirow{5}{*}{\textbf{AlexNet}}                  & \textbf{HBR}                                  & \textbf{100}  & 0             & 0             & 0             & 0             \\
                                                              &                                                    & \textbf{HK}                                   & 0.3           & \textbf{91.8} & 3.9           & 0.5           & 3.5           \\
                                                              &                                                    & \textbf{HP}                                   & 0.6           & 0.8           & \textbf{98.3} & 0             & 0.3           \\
                                                              &                                                    & \textbf{HG}                                   & 0.2           & 1.2           & 8.5           & \textbf{89.3} & 0.8           \\
                                                              &                                                    & \textbf{HW}                                   & 0             & 1.4           & 0             & 0             & \textbf{98.6} \\ \cline{2-8} 
                                                              & \multirow{5}{*}{\textbf{GoogLeNet}}                & \textbf{HBR}                                  & \textbf{99.4} & 0             & 0             & 0.6           & 0             \\
                                                              &                                                    & \textbf{HK}                                   & 0.2           & \textbf{88.6} & 1.5           & 1             & 8.7           \\
                                                              &                                                    & \textbf{HP}                                   & 0.3           & 4.9           & \textbf{93.6} & 1.1           & 0.1           \\
                                                              &                                                    & \textbf{HG}                                   & 0             & 1.2           & 0.3           & \textbf{98.5} & 0             \\
                                                              &                                                    & \textbf{HW}                                   & 0             & 0.5           & 0             & 0.1           & \textbf{99.4} \\ \cline{2-8} 
                                                              & \multirow{5}{*}{\textbf{ResNet18}}                 & \textbf{HBR}                                  & \textbf{100}  & 0             & 0             & 0             & 0             \\
                                                              &                                                    & \textbf{HK}                                   & 0             & \textbf{93.6} & 1.7           & 1             & 3.7           \\
                                                              &                                                    & \textbf{HP}                                   & 0             & 0.6           & \textbf{97.6} & 1.7           & 0.1           \\
                                                              &                                                    & \textbf{HG}                                   & 0             & 0.5           & 0.1           & \textbf{99.4} & 0             \\
                                                              &                                                    & \textbf{HW}                                   & 0             & 0.2           & 0             & 0             & \textbf{99.8} \\ \hline
    \end{tabular}
    \end{table*}
    \item \textbf{Case 1c: Train using data from multiple bistatic angles:} Finally, we train the algorithms with a simulation dataset comprising micro-Doppler data captured at different bistatic angles. It is a more practical scenario that one can encounter. Here, we observe the performance to be lower compared to the previous two cases. This is because the micro-Doppler signatures are captured in a forward scatter geometry, resulting in micro-Dopplers being centered mostly around 0 Doppler for almost all the motion classes. This leads to a significant reduction in the classification accuracies. We present the classification results for this case in TABLE \ref{table:bi_vary_class_sim}. We note that HW's performance is good across all the algorithms ($\geq $ 90). Compared to the previous case, there is a drop of $\approx$ 3\% in classification accuracy for AlexNet, GoogLeNet, and DCNN. ResNet seems to work best amongst all classifiers with an average classification accuracy of up to 95.8\% (2\% reduction compared to the previous case).   
    
    \begin{table*}[htbp]
    \centering
    \caption{Classification accuracies for a simulation database (captured under varying bistatic circular configurations)}
    \label{table:bi_vary_class_sim}
    \begin{tabular}{|c|c|c|c|c|c|c|c|}
    \hline
    \multirow{2}{*}{\textbf{Target Class/ Algorithm}} & \multirow{2}{*}{\textbf{\begin{tabular}[c]{@{}c@{}}Handpicked\\  Features\end{tabular}}} & \multirow{2}{*}{\textbf{\begin{tabular}[c]{@{}c@{}}CVD \\ Features\end{tabular}}} & \multirow{2}{*}{\textbf{\begin{tabular}[c]{@{}c@{}}Sparse\\  Features\end{tabular}}} & \multicolumn{4}{c|}{\textbf{Neural Networks}}                             \\ \cline{5-8} 
                                                      &                                                                                          &                                                                                   &                                                                                      & \textbf{DCNN} & \textbf{AlexNet} & \textbf{GoogLeNet} & \textbf{ResNet18} \\ \hline
    HBR                                               & 93.6                                                                                     & 91.8                                                                              & 99.1                                                                                 & 80.2          & 82.6             & 87.2               & 91.2              \\ \hline
    HK                                                & 17.2                                                                                     & 35.5                                                                              & 81.8                                                                                 & 86.9          & 83.9             & 85.8               & 91.6              \\ \hline
    HP                                                & 21.2                                                                                     & 88.7                                                                              & 89.7                                                                                 & 97            & 95.9             & 91.3               & 99.5              \\ \hline
    HG                                                & 64.4                                                                            & 60.3                                                                     & 89.3                                                                       & 94.8 & 97.9    & 93.6      & 95.1     \\ \hline
    HW                                                & 90.9                                                                                     & 73.8                                                                              & 95.8                                                                                 & 98.6          & 99.2             & 100                & 99.4              \\ \hline
    \textbf{Overall Accuracy (\%)}        & \textbf{57.46}                                                                                    & \textbf{70.2}                                                                              & \textbf{91.14}                                                                                & \textbf{91.5}          & \textbf{91.9}             & \textbf{91.58}              & \textbf{95.79}             \\ \hline
    \end{tabular}
    \end{table*}
\end{itemize}

\subsection{Performance Under Noisy Conditions}
The results presented so far have been computed for high signal to noise ratios ($SNR \approx 30$ dB). Most realistic scenarios have environmental factors that significantly affect the resulting SNRs. Therefore, to assess SNR's impact on the classification performances, we introduced additive Gaussian noise to achieve varying levels of SNR (-2 to 10dB) to the simulated micro-Doppler signatures. The simulation database used for this study is captured for a fixed $0^{\circ}$ aspect angle of the target. We carried out ten repetitive trials using randomly selected training and test noisy micro-Doppler signatures.

\begin{figure}[t]
\centering
\includegraphics[scale=0.4]{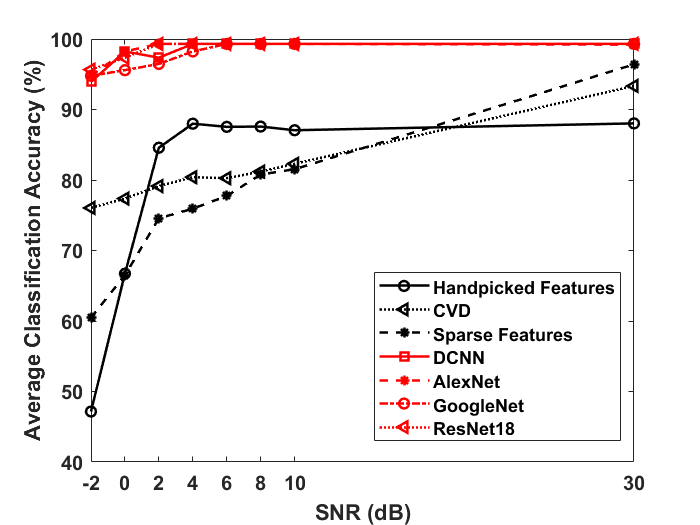}
\caption{Average classification accuracies as a function of SNR}
\label{fig:Sim_SNR}
\end{figure}

Fig. \ref{fig:Sim_SNR} shows the variation of average classification performances of different algorithms as a function of SNR. As the noise level is increased, we observe the expected drop in different algorithms' overall performance. However, the drop in performance is higher for classical learning-based methods (represented in black) than deep learning models (represented in red). The results show that the deep networks can extract features that are more robust to noise  and therefore should be the 'classifier of choice' in high-noise scenarios. 
\section{Measurements}
\label{sec:measurements}
In this section, we do a more thorough evaluation of the algorithms' performance in real-world scenarios.
\subsection{Measurement Data Collection}
We deployed the PWR measurement setup and the motion capture system to simultaneously gather the simulation animation data and the measurement PWR data from the same motion classes HBR, HK, HP, HG, and HW. 
The PWR system was implemented using software-defined radio (SDR) as presented in our previous work \cite{li2020passive}, whereas two USRP-2921 \cite{USRP2921} were used as RF front-end. 
The collected raw WiFi signal was then transferred to a computer and processed with cross ambiguity function. 
\begin{figure}[t]
\centering
\includegraphics[scale=0.50]{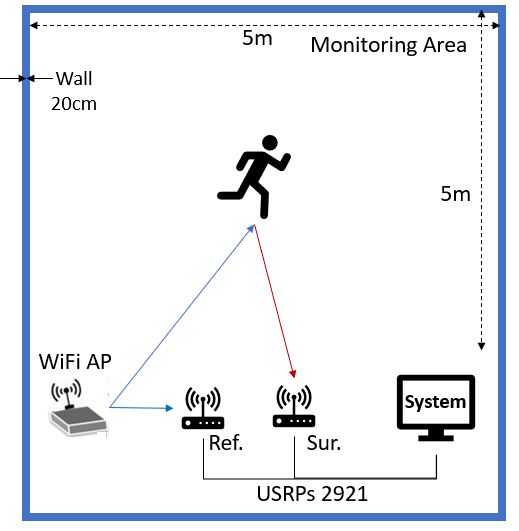}
\caption{Measurement setup comprising of two USRP 2921, a WiFi AP and a signal processing unit. USRP connected to reference antenna captures the direct transmissions from WiFi AP. Second USRP is used to capture the signals reflected off the targets present in the monitoring area.}
\label{fig:meas_setup}
\end{figure}
The measurement setup is depicted in Fig.\ref{fig:meas_setup}. We used a Raspberry Pi to attempt handshakes with a WiFi access point ('AP') to generate WiFi transmissions. By constantly transmitting probe requests, the WiFi AP continuously emitted probe response signals. We captured these transmissions at two antennas- a reference antennas and a surveillance antenna connected to two USRPs. The reference channel receives direct transmissions from the WiFi AP, while the surveillance channel gathers signals reflected off the targets moving in the same propagation channel. 

\begin{figure*}[t]
\centering
\includegraphics[scale=0.60]{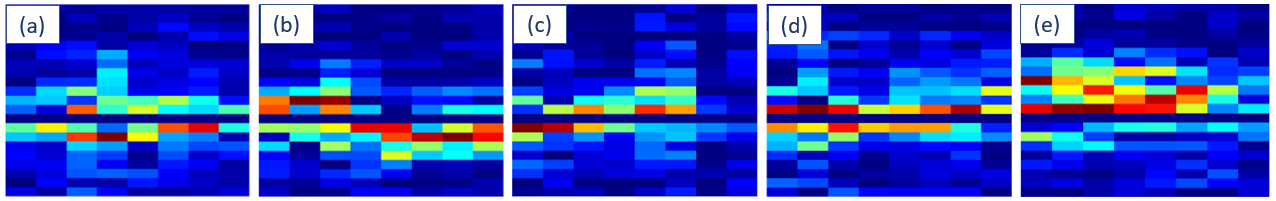}
\caption{Real micro-Doppler signatures for a human undergoing (a) a body rotation motion, (b) kicking motion, (c) punching motion, (d) grabbing an object motion and (e) a human walking in the direction of the monostatic configuration of PWR radar.}
\label{fig:meas_spec}
\end{figure*}

Fig.\ref{fig:meas_spec}(a)-(e), shows measured spectrograms corresponding to the five motion classes. We can see from the spectrograms that four target classes (HBR, HK, HP, and HG), have a periodic motion with positive and negative Doppler frequencies. HW spectrogram, shown in Fig.\ref{fig:meas_spec}(e), can be discerned from other motion classes as it has a Doppler shift due to bulk body motion and additional micro-Dopplers. However, due to several environmental factors such as multipath, shadowing, path-loss, the measured spectrograms are noisy. 

TABLE \ref{table:meas_dataset_des} summarizes our entire measurement data. 
\begin{table*}[]
\centering
\caption{Measurement Dataset Description}
\label{table:meas_dataset_des}
\begin{tabular}{|c|c|c|c|l|l|l|l|}
\hline
\textbf{Target Class} & \textbf{\begin{tabular}[c]{@{}c@{}}Number of Measurement \\ Files\end{tabular}} & \textbf{\begin{tabular}[c]{@{}c@{}}Duration of Data\\ (in each file)\end{tabular}} & \textbf{\begin{tabular}[c]{@{}c@{}}Number of Spectrograms  (1.5sec each)\\ In Monostatic PWR Configuration Scenario\\ (Overlapping Window of 0.5sec)\end{tabular}} \\ \hline
HBR                   & 20                                                                              & 4.5 sec                                                                            & 180                                                                                                                                                               \\ \hline
HK                    & 20                                                                              & 4.5 sec                                                                            & 180                                                                                                                                                               \\ \hline
HP                    & 20                                                                              & 4.5 sec                                                                            & 180                                                                                                                                                                \\ \hline
HG                    & 20                                                                              & 4.5 sec                                                                            & 180                                                                                                                                                                \\ \hline
HW                    & 20                                                                              & 4.5 sec                                                                            & 180                                                                                                                                                               \\ \hline
\multicolumn{3}{|c|}{\textbf{Total Data in Each Configuration}}                                                                                                                              & \textbf{900}                                                                                                                                                       \\ \hline
\end{tabular}
\end{table*}
\subsection{Measurement Results and Analyses}
We validated the performance of different algorithms across a common measurement dataset, and the resulting accuracies are reported in TABLE \ref{table:meas_class_mono}. Since the real spectrograms subsume environment effects such as noise, multipath, and shadowing, we expect the performances to be lower than that using simulations. We can already see from the results that the handpicked features and CVD features completely fail to classify motion classes in 60\% of the cases. Because the real spectrograms have noise and multipath components, the spectrograms can no longer be well represented through sparse features. Therefore, the average classification accuracy sparsity-based algorithm is reduced to only 66.6\% . Hence, we believe that the classical machine learning algorithms are not suited to deal with a great deal of diversity in the radar data due to various environmental factors such as noise and multipath. 

On the other hand, the neural network's results indicate that deep networks are more robust to environmental factors such as noise and can perform well even in diverse operating conditions. We were able to obtain classification accuracies $\approx$ 90\% for all the neural networks. The ResNet performs best with an average classification accuracy of 92.6\%.
\begin{table*}[]
\centering
\caption{Classification accuracies of different algorithms using measurement data (captured using monostatic configuration of PWR radar)}
\label{table:meas_class_mono}
\begin{tabular}{|c|c|c|c|c|c|c|c|}
\hline
\multirow{2}{*}{\textbf{Target Class/ Algorithm}} & \multirow{2}{*}{\textbf{\begin{tabular}[c]{@{}c@{}}Handpicked\\  Features\end{tabular}}} & \multirow{2}{*}{\textbf{\begin{tabular}[c]{@{}c@{}}CVD \\ Features\end{tabular}}} & \multirow{2}{*}{\textbf{\begin{tabular}[c]{@{}c@{}}Sparse\\  Features\end{tabular}}} & \multicolumn{4}{c|}{\textbf{Neural Networks}}                             \\ \cline{5-8} 
                                                  &                                                                                          &                                                                                   &                                                                                      & \textbf{DCNN} & \textbf{AlexNet} & \textbf{GoogLeNet} & \textbf{ResNet18} \\ \hline
HBR                                               & 31.3                                                                                     & 45.5                                                                              & 48.7                                                                                 & 100           & 81.5             & 85.2               & 88.9              \\ \hline
HK                                                & 3.1                                                                                      & 64.7                                                                              & 90.9                                                                                 & 85.2          & 85.2             & 85.2               & 100               \\ \hline
HP                                                & 73.7                                                                                     & 85.3                                                                              & 71.9                                                                                 & 74.1          & 96.8             & 100                & 85.2              \\ \hline
HG                                                & 21.1                                                                            & 12.5                                                                     & 39.5                                                                        & 92.6 & 93.1    & 77.8      & 92.6     \\ \hline
HW                                                & 77.5                                                                                     & 9.7                                                                               & 81.8                                                                                 & 96.3          & 96.3             & 96.3               & 96.3              \\ \hline
\textbf{Overall Accuracy (\%)}        & \textbf{41.3}                                                                                    & \textbf{43.5}                                                                             & \textbf{66.6}                                                                                & \textbf{89.6}         & \textbf{90.6}            & \textbf{88.9}               & \textbf{92.6}             \\ \hline
\end{tabular}
\end{table*}

\section{Conclusion}
\label{sec:conclusion}
In this work, we first presented a bespoke simulator that can simulate human micro-Doppler radar returns as a function of a diverse set of target parameters, radar parameters, and radar signal processing parameters. We used three different simulation databases with different parameter variations to study and evaluate machine learning and deep learning algorithms' classification performances in more complex scenarios. We achieved average classification accuracies $\geq$ 90\% for almost all the deep learning models. We also demonstrated the performance analysis of different algorithms as a function of SNR.  The classification was challenging due to the close similarity of motion classes considered in the study. The results suggest that the deeper networks are less affected by the noise than classical methods, thus making them an ideal candidate for real-world scenarios comprising various environmental factors such as noise, and multipath. To validate the performance of the algorithms, radar data was captured simultaneously with the motion capture animation data to maintain consistency across both simulations and measurements. We were able to get good classification accuracies (up to 90\%) even with the real dataset compared to those obtained from simulations. 

The study demonstrates the feasibility of passive WiFi sensing for activity recognition applications in real-world scenarios. Current experiments presented in the paper are limited due to the small number of participants and the diversity of experimental scenarios (considered only line-of-sight conditions). Therefore, we plan to extend our simulation framework to incorporate more participants and more realistic indoor through-wall scenarios, including environmental factors such as multipath and shadowing. We also plan to add more motion categories like falling, hand gestures, bending down to our existing simulator. Interested researchers can download the simulator from \url{https://uwsl.co.uk/}. We believe the simulator will pave the way to benchmark future algorithms and generate large volumes of simulation data.   
\section*{Acknowledgments}
This work is part of the OPERA project funded by the UK Engineering and Physical Sciences Research Council (EPSRC), Grant No: EP/R018677/1. 
\bibliographystyle{IEEEtran}
\bibliography{References}
\vspace{-1cm}
\end{document}